\documentclass[utf8]{FrontiersinHarvard} 

\usepackage{url,hyperref,lineno,microtype,subcaption}
\usepackage[onehalfspacing]{setspace}

\usepackage{amsmath}
\usepackage{amsfonts}
\usepackage{amssymb,upgreek,natbib}
\usepackage{bm}
\usepackage[utf8]{inputenc}
\usepackage{multirow}
\usepackage{graphicx}
\usepackage{makecell}
    \usepackage{ulem} 
    \usepackage{cancel} 

\newcommand{\e}{\mathrm{e}}  
\newcommand{\dd}{\mathrm{d}}  
\newcommand{\ma}{_\text{m}}  
\renewcommand{\vec}{\bm} 

\newcommand{\const}{\text{const}} 
\newcommand\deriv[2]{\displaystyle\frac{\partial #1}{\partial #2}}

\newcommand{\CR}{_\text{CR}}  
\newcommand{\SN}{_\text{SN}}  
\newcommand{\SFR}{\dot{M}}     

\newcommand{\z}{\tilde{z}} 
\newcommand{\erf}{\operatorname{erf}}

\newcommand{\kpc}{\,\text{kpc}}  
\newcommand{\mkG}{\,\upmu{\rm G}}  
\newcommand{\cm}{\,\text{cm}}
\newcommand{\m}{\,\text{m}}
\newcommand{\Hz}{\,\text{Hz}}
\newcommand{\GHz}{\,\text{GHz}}
\newcommand{\MHz}{\,\text{MHz}}

\newcommand{\kms}{\,\text{km}\s^{-1}}

\newcommand{\s}{\,\text{s}}    
\newcommand{\yr}{\,\text{yr}}    

\newcommand{\erg}{\,\text{erg}}    
\newcommand{\eV}{\,\text{eV}}    
\newcommand{\MeV}{\,\text{MeV}}  
\newcommand{\GeV}{\,\text{GeV}}  

\definecolor{mypurple}{rgb}{0.7,0.3,0.8}
\newcommand{\as}[1]{#1}
\newcommand{\ass}[1]{}

\def\keyFont{\fontsize{8}{11}\helveticabold }
\def\firstAuthorLast{Shukurov and Jose} 
\def\Authors{Anvar Shukurov\,$^{1,*}$ and Charles Jose\,$^{2,*}$}
\def\Address{$^{1}$School of Mathematics, Statistics and Physics, Newcastle University, UK\\
$^{2}$Department of Physics, CUSAT, Cochin, India}
\def\corrAuthor{A.~Shukurov, School of Mathematics, Statistics and Physics, Newcastle University, Newcastle upon Tyne, NE1~7RU, UK}

\def\corrEmail{anvar.shukurov@ncl.ac.uk}

\begin{document}
\onecolumn
\firstpage{1}

\title[Cosmic Ray Electrons]{The Distribution of Cosmic Ray Electrons in Star-Forming Galaxies}

\author[\firstAuthorLast ]{\Authors} 
\address{} 
\correspondance{} 

\extraAuth{C. Jose, Department of Physics, CUSAT, Cochin, 682022, India \\ charles.jose@cusat.ac.in}

\maketitle

\begin{abstract}
\section{}
We derive explicit, algebraic expressions for the steady-state number density of cosmic ray electrons as a function of position and energy using Green's function of the diffusion equation with energy losses for an axisymmetric distributions of the particle sources in the galactocentric radius $r$ and distance to the mid-plane $z$. The solution is obtained for a Gaussian distribution of the particle sources in $r$ and $z$ but we show that it can be used for an arbitrary spatial distribution of the sources. The accuracy of our results is about 10\% or better in \ass{a} wide \ass{range} \as{ranges} of $r$, \ass{and} $z$ and particle energies. These solutions can be used in the interpretation of radio astronomical observations of galaxies, particularly in the  studies of the radio luminosities for large galaxy samples, and represent a physically justifiable and efficient alternative to the assumption of the energy equipartition between cosmic rays and interstellar magnetic fields.

\tiny
 \keyFont{ \section{Keywords:} cosmic ray electrons, advection-diffusion approximation, synchrotron emission, star-forming galaxies, galaxy formation, radio luminosity function} 
\end{abstract}

\section{Introduction}
The spectrum and spatial distribution of the synchrotron emission of cosmic ray (CR) electrons (CRE), observable in the radio range, carries abundant information about the interstellar medium (ISM) of galaxies. The spectrum of the radio emission is controlled by the energy spectrum of the particles which depends, together with their spatial distribution, on the distribution of the cosmic ray sources and the particle propagation in the ISM, apart from the particle acceleration mechanisms. Energy losses to the synchrotron emission and inverse Compton scattering strongly affect both the energy spectrum and the propagation of the relativistic electrons.

For our purposes, CR propagation can be described using the fluid (advection-diffusion) approximation \citep{GS64,BBGDP90,S02}. General CR propagation codes are available \citep[e.g.,][]{S+10,E+08} apart from numerous simulations of specific objects \citep[see][for reviews]{AB18,HSG21,H25}. However, the interest in explicit, analytic solutions of the CR transport equations persists. Such solutions are required for the interpretation of radio astronomical observations of well resolved galaxies \citep[e.g.,][]{S+2023,I+24} when  multi-dimensional simulations are impractical and numerical solutions of a simplified, one-dimensional transport equation are employed \citep{MFBMS16,H+18,H2021} as an alternative to semi-analytic solutions \citep[e.g.,][]{S77,RS19,RS24}. Another emerging application area for such solutions is the interpretation of the radio luminosity functions of statistically large samples of galaxies which are becoming a powerful diagnostic of the galaxy formation theory. In such applications \citep[e.g.,][and references therein; Ghosh et al., in preparation]{Schober+2023, JCSSRB24, Hansen+2024, Yoon2024, Th+26}, the simplicity and explicit form of the particle spectrum and spatial distribution are often more important than the accuracy and generality of the CRE propagation model. In such cases, only an explicit form for the spatial distribution of CRE is a suitable option since solving CR transport equations for each object is computationally prohibitive. Even the usefulness of an expression for the particle number density and energy in the form of a multiple integral is highly problematic in this case. Most applications of this kind derive the CR energy density using the assumption of energy equipartition with interstellar magnetic fields. This assumption is questionable and a better, physically justifiable model is required \citep{SB19} based on the CRE propagation theory. In this paper, we develop an approximate explicit solution of the diffusion equation for the CRE distribution and energy spectrum in a star-forming disc galaxy. Unavoidably, such a solution involves simplifications which, however, allow us to obtain flexible, accurate and simple general results.

The relation of the intensity of CRE sources   to the galactic parameters is discussed in Section~\ref{IRCR}. Using Green's function of \citet{S59} \citep[see also][]{BBGDP90,AAV95}, we derive in Section~\ref{DSRE} explicit, algebraic expressions for the energy spectra and axially symmetric spatial distributions of CR electrons assuming Gaussian profiles of the CR source intensity along the galactocentric distance and across the disc. The accuracy of the results in discussed in Section~\ref{QoA}, and Section~\ref{app:mid_energy} presents further refinements of the approximate solutions. Their generalisation to an arbitrary spatial distribution of CRE sources is the subject of Section~\ref{ARDCRS}. Implementation to galaxies with a strong spatial variation of the magnetic field strength and the energy density of the stellar radiation field is discussed in Section~\ref{AoA} which also summarizes the results.

\section{Injection rate of cosmic rays}\label{IRCR}
Supernovae are the main source of CR in star-forming galaxies \citep{BBGDP90,L94,S02,B13}. About $\epsilon\CR=0.03\text{--}0.1$ of the energy of a supernova explosion $E\SN=10^{51}\erg$ is converted into the energy of the relativistic particles \citep[equation 3 of \citealt{B13}; section~2.3 of][]{BSY14}. The value of $\epsilon\CR$ only weakly depends on the slope of the injection energy spectrum $s_0$ \citep{B13}. The ratio of the number densities of relativistic electrons and protons depends on the ratio of their rest masses $m_\e$ and $m_\text{p}$ as $\delta_\e\simeq (m_\e/m_\text{p})^{(s_0-1)/2}$ \citep[\citealt{Bell78b}; section~19.4 of][see however section~3.8 of \citealt{SMP07}]{S02}. For $s_0 =2.2$, this implies $\delta_\e\simeq 10^{-2}$ in agreement with observations. 

The fraction of stars that evolve to supernovae (stellar masses $10<M/M_\odot<40$) is $\delta\SN=8\times10^{-3}$ for the initial stellar mass function of \citet{K01,K08}, and the corresponding average stellar mass is $M_\star=0.85\,M_\odot$. For the global star formation rate $\SFR$, the supernova frequency follows as $\nu\SN=\delta\SN \SFR/M_\star$. For the Milky Way, where $\SFR=3\,M_\odot\yr^{-1}$ \citep{R91}, this leads to $\nu\SN=0.028\yr^{-1}$.

The energy supply rate to CR (galactic CR luminosity) follows as
\begin{equation}\label{WCR}
	W\CR =\delta\SN \epsilon\CR \frac{\SFR}{M_\star}E\SN
	\simeq 10^{40}\erg\s^{-1} \left(\frac{\epsilon\CR}{0.03}\right) \left(\frac{\SFR}{1\,M_\odot\yr^{-1}}\right),
\end{equation}
and relativistic electrons receive the fraction $\delta_\e$ of this amount. 

The synchrotron emission of a relativistic electron of an energy $E$ in a magnetic field of a strength $B$ has a maximum near the frequency \ass{(about 0.3 of the maximum emission of a single electron)} \citep{BBGDP90,L94}:
\begin{equation}\label{num}
	\nu\ma = 4.8\times10^6 \Hz \left(\frac{B}{1\mkG} \right)
	\left(\frac{E}{1\GeV}\right)^2\,.
\end{equation}
We assume that all the particle energy is radiated away at this frequency \citep[detailed discussion of the single-particle spectrum can be found, e.g., in section~4 of][]{L94}.

The radio luminosity functions of galaxies  are often obtained at the rest-frame frequencies of $1.42\GHz$, $408\MHz$ and $150 \MHz$ (the wavelengths $\lambda=21\cm$, $73.5\cm$ and $2\m$, respectively). The energy of the electrons emitting at a given frequency $\nu\ma$ is given by
\begin{equation}\label{Em}
	E\ma = \as{\xcancel{8} 7.7}\GeV \left(\frac{5\mkG}{B}\right)^{1/2} \left(\frac{\nu\ma}{1.42\GHz}\right)^{1/2}\,.
\end{equation}

\section{Propagation of relativistic electrons}\label{DSRE}

In the diffusion--advection approximation, the distribution of CRE is governed by \citep[section~5.3 of][]{BBGDP90}
\begin{equation}\label{ADeq}
	\deriv{N}{t}= \nabla\cdot[D\nabla N))- \vec{u}N]-\deriv{}{E}[b(E)N] -\frac{N}{T} + Q(\vec{x},t,E)\,,
\end{equation}
where $N(\vec{x},t,E)$ is the number density of the particles per unit energy interval ($[N]= \text{cm}^{-3}\GeV^{-1}$), $D$ is the diffusivity (we assume that the diffusion is isotropic), $\vec{u}$ is the advection velocity, $b(E)\equiv \dd E/\dd t$ is the particle energy loss rate, $T$ is the time scale of particle loss from the system and $Q(\vec{x},t,E)$ is the density of the particle sources per unit energy interval  ($[Q]= \text{cm}^{-3}\s^{-1}\GeV^{-1}$).

The energy loss rate of a CR electron to the synchrotron emission and inverse Compton scattering off photons with the energy density $w_\text{ph}$ is given by
\begin{equation}\label{Eloss}
	b(E)= -\beta E^{2}\,,
	\qquad
	\beta = \frac{1}{1.3\times10^{10}\yr\GeV} \left[\left(\frac{B}{1\mkG}\right)^2 + \frac{w_\text{ph}}{0.025\eV\cm^{-3}} \right].
\end{equation}
The energy loss rate depends on the galactic magnetic field strength $B$ which varies across the galaxy while the energy density of photons $w_\text{ph}$ depends on the redshift in the case of the cosmic microwave background (CMB) ($w_\text{ph}=4.2\times10^{-13}(1+\z)^4\erg\cm^{-3} = 0.26 (1+\z)^4 \eV \cm^{-3}$ , where $\z$ is the redshift) and also on the position within the galaxy if its radiation field is included. Although the electron energy losses due to the stellar radiation can be significant at $\z \lesssim 1$, especially in the central parts of galaxies, they might be neglected at higher redshifts in comparison with the CMB losses (Section~\ref{AoA}). 
\as{We neglect the Klein--Nishina effect which reduces the rate of the radiative energy losses of electrons at $E\gtrsim 50\GeV$ \citep[e.g., section~4.2.1 of][]{S02}. We also neglect any bremsstrahlung and ionization energy losses.}

Following \citet{S59} and \citet[][section~5.4]{BBGDP90}, we consider the following axisymmetric distribution of the cosmic rays sources:
\begin{equation} \label{Q} 
	Q(\vec{x},E) = \frac{KE^{-s_0}}{\pi^{3/2} R^2 h} \exp\left(-\frac{r^2}{R^2}-\frac{z^2}{h^2}\right)\,, 
\end{equation}
in terms of the cylindrical coordinates $(r,\phi,z)$, with the radial and vertical length scales $R$ and $h$, the injection spectral index $s_0$ and a constant $K$. For simplicity, we adopt $s_0=2$ wherever possible, and then $[K]=\text{GeV}\s^{-1}$. The total energy injection rate follows as
\begin{equation}
	\int \dd^3\vec{x} \int dE \, E Q(\vec{x}, E) = W_{\text{CR}}.
\end{equation}
For the normalisation adopted in equation~\eqref{Q}, this reduces to
\begin{equation}
	K\int_{E_{\text{min}}}^{E_{\text{max}}} \dd E \, E^{1 - s_0} = W_{\text{CR}}\,,
\end{equation}
and then
\begin{equation}\label{KWcr}
	K =
	\begin{cases}
		\dfrac{W_{\text{CR}}}{\ln \left(E_{\text{max}}/E_{\text{min}} \right)} & \text{if } s_0 = 2\,, \\[1em]
		\dfrac{(s_0 - 2)W_{\text{CR}}}{ E_{\text{min}}^{2 - s_0} - E_{\text{max}}^{2 - s_0}} & \text{if } s_0 > 2\,.
	\end{cases}
\end{equation}
We adopt $E_{\text{min}} =0.5\MeV$, close to the electron rest mass and energy at which the CRE energy spectrum flattens \citep{SMP07}. For numerical estimates  and the analytic expressions for $N(\vec{x},E)$ in Sections~\ref{BEEE} and \ref{app:mid_energy}, we assume $s_0=2$, with the upper energy limit taken to be $E_{\max} = 10^8\GeV$, which is  large enough to have a negligible impact on the results.

We assume that the density of the particle sources is independent of time and derive the steady-state spatial distribution and energy spectrum of CRE in the diffusion approximation. Our results remain applicable to evolving galaxies as long as the characteristic time of the development of the steady states in the particle distribution (depending on the electron diffusion, advection and energy loss time scales) is much shorter than the characteristic times of the galactic evolution. For the CR diffusivity $D=3\times10^{28}\cm^2\s^{-1}$ \citep{SMP07}, the diffusion time over the distance of $L=1\kpc$ is as short as $L^2/D=10^7\yr$. 

Steady-state solutions of the transport equation \eqref{ADeq} discussed below are obtained under the following simplifying assumptions. The energy loss rate \eqref{Eloss} is assumed to be position-independent. This assumption is fully acceptable for the inverse Compton scattering off the CMB photons but not for losses to the synchrotron and stellar radiation. We discuss in Section~\ref{AoA} how the spatial variation of the galactic magnetic and radiation fields can be accounted for. We assume that the CR diffusion is isotropic and neglect the dependence of the diffusivity on position and energy. 

\subsection{The diffusion approximation}
Green's function of equation~\eqref{ADeq}, its solution for $Q=\updelta(\vec{x}-\vec{x}_0)\, \updelta(t-t_0)\, \updelta(E-E_0)$, in infinite space and for $\vec{u}=0$ (the diffusion approximation) is given by \citep{S59}
\begin{equation}\label{Green}
	G( \vec{x}, t, E ; \vec{x}_0, t_0, E_0)=\frac{\exp \left[-\tau / T-\left(\vec{x}-\vec{x}_{0}\right)^{2} / (4 \Lambda^{2})\right]}{|b(E)|\left(4 \pi \Lambda^{2}\right)^{3 / 2}} \delta\left(t-t_{0}-\tau\right),
\end{equation}
where 
\as{the average path length of an electron with an initial energy $E_0$ and a final energy $E$ is given by}
\begin{equation} \label{lambda}
	\Lambda(E, E_0)=\left[\int_{E_0}^E \frac{D(E')\,\dd E'}{b(E')}\right]^{1/2}
\end{equation}
\ass{is the average path length of an electron with an initial energy $E_0$ and a final energy $E$,} and
\begin{equation}\label{tau}
	\tau(E, E_0)=\int_{E_0}^{E} \frac{\dd E'}{b(E')} = \frac{1}{\beta}\left(\frac{1}{E}-\frac{1}{E_0}\right)
\end{equation}
is the time scale of the energy loss from $E$ to $E_0$.

The CRE number density per unit energy interval is given by
\begin{equation}\label{NQ}
	N(\vec{x}, t, E)=\int_{V} \dd^3 \vec{x}_0 \int_{-\infty}^t \dd t_0 \int_0^\infty \dd E_{0} \, Q\left(\vec{x}_0, t_0, E_0\right) G\left(\vec{x}, t, E ; \vec{x}_0, t_0, E_0\right)\,,
\end{equation}
where the volume integral extends over the infinite space.
The integral over $t_0$ leads to the step function which differs from zero only if $\tau>0$, i.e., $E<E_0$ \citep{S59}. Therefore, the integral over $E_0$ extends over the range $E<E_0<\infty$. For the electrons, any losses are negligible in comparison with the synchrotron and inverse Compton scattering, so that $\exp(-\tau/T)\approx1$.

The energy loss rate $b(E)$ and, consequently, $\tau$ and $\Lambda$, depend on position, in particular because $B$ is a function of $\vec{r}$. In order to evaluate the integral, we neglect this dependence and replace $B^2$ by its mean value in applications (see Section~\ref{AoA}). We also assume that $D=\const$ ($=3\times10^{28}\cm^2\s^{-1}$), and then the mean free path of a relativistic electron reduces to
\begin{equation}\label{mfp}
	\Lambda=(D\tau)^{1/2}\,.
\end{equation}

Neglecting for simplicity the  contribution of the stellar radiation to the inverse Compton scattering, 
the half-energy loss time of an electron and the corresponding mean free path at a redshift $\z$ are 
\begin{equation}\label{h-time}
\begin{split}
	\tau_{1/2}& =\frac{1}{\beta E_0} \simeq \frac{1.3\times10^9\yr}{(1+\z)^4+ (B/3.2\mkG)^2} \, \left(\frac{E_0}{1\GeV}\right)^{-1}\,,\\
    \Lambda_{1/2}&=(D\tau_{1/2})^{1/2}\simeq \frac{11\kpc}{[(1+\z)^4+ (B/3.2\mkG)^2]^{1/2}} \, \left(\frac{E_0}{1\GeV}\right)^{-1/2}\,.
    \end{split}
\end{equation}

The integrals over the spatial variables in equation~\eqref{NQ} with $Q_0$ of the form \eqref{Q} are convenient to evaluate in Cartesian coordinates $(x,y,z)$ \citep[section 5.4 of][]{BBGDP90} where they reduce to $\int_{-\infty}^\infty \exp(-\xi^2-a\xi)\,\dd\xi=\sqrt{\pi}\exp(a^2/4)$, leading to ($r^2=x^2+y^2$) 
\begin{equation}\label{NQC}
	N(\vec{x}, E)=\frac{K}{\pi^{3/2} {|b(E)|}}
	\int_E^\infty  \frac{\dd E_0}{E_0^{s_0}} 
	\frac{\displaystyle\exp\left(-\frac{r^2}{R^2+4\Lambda^2}-\frac{z^2}{h^2+4\Lambda^2} \right)}{(R^2+4\Lambda^2)(h^2+4\Lambda^2)^{1/2}}\,.
\end{equation}
\as{This relation is valid for any value of the injection spectral index $s_0$ and when the diffusion coefficient depends on the particle energy. However, the integral over $E$ can be evaluated only for $s_0=2$ and $D=\const$ to obtain explicit expressions for $N(\vec{x}, E)$ presented below. We discuss in Section~\ref{DC} the dependence of $D$ on the particle energy.}

\subsection{The energy spectrum and spatial distribution of CRE }\label{BEEE}
The energy spectrum of the particles in the diffusion approximation has different forms in three energy ranges controlled by the relation between the mean free path of the particles and the vertical and radial sizes of the system, i.e., by the relation between $4\Lambda^2=4D\tau$ and $h^2$ and $R^2$.
Numerical estimates presented below are obtained for $h=0.1\kpc$, $R= 10\kpc$ and $D=3\times10^{28}\cm^2\s^{-1}$.
The expressions for $N(\vec{x},E)$ are derived in this section assuming that $s_0 = 2$ and  $E\ll E_0$ (so that  that $\tau\approx (\beta E)^{-1}$). Refinements based on a more careful consideration of the particle propagation length as a function of $E_0$ are presented in Section~\ref{app:mid_energy}. 

\subsubsection{High energies}\label{I1}
For $4\Lambda^2\ll h^2\ll R^2$, i.e.,
\begin{equation}\label{hE}
	E\gg \frac{4D}{\beta h^2} \simeq \frac{5.2\times10^4\GeV}{(1+\z)^4+ \left(B/3.2\mkG\right)^2}\,,\qquad \Lambda_{1/2}\ll h/2\,,
\end{equation}
we have $\exp[-z^2/(h^2+4\Lambda^2)]\approx \exp(-z^2/h^2)$ and $\exp[-r^2/(R^2+4\Lambda^2)]\approx \exp(-r^2/R^2)$. The only energy-dependent term in the integrand of equation~\eqref{NQC} is then $E_0^{-2}$, leading to
\begin{equation}\label{NrE1}
	N(\vec{x},E)\approx \frac{K}{\pi^{3/2}\beta R^2h} E^{-3}\exp\left(-\frac{r^2}{R^2}-\frac{z^2}{h^2} \right).
\end{equation}
These particles do not propagate far from their sources before they lose their energy and their spectral index is equal to $-s_0-1=-3$ for $s_0=2$.

Since $\exp\left[-z^2/(h^2+4\Lambda^2)\right] 
\approx \exp\left(-z^2/h^2\right)\left(1+4\Lambda^2 z^2/h^4\right)$, this approximation for $N(\vec{r},E)$, where $4\Lambda^2z^2/h^4$ and higher-order terms are neglected, is valid at 
\begin{equation}\label{Zmax}
	|z_{1/2}| \lesssim h^2/(2\Lambda_{1/2})\simeq h\,,
    \qquad
    r_{1/2}\lesssim R^2/(2\Lambda_{1/2})\simeq R^2/h\,,
\end{equation}
in terms of the half-energy mean free path $\Lambda_{1/2}$ and the corresponding distances $z_{1/2}$ and $r_{1/2}$ evaluated here at the extreme vales of $\Lambda_{1/2}$ in equation~\eqref{hE}.

\subsubsection{Intermediate energies}\label{I2}
For $h^2\ll 4\Lambda^2 \ll R^2$, i.e., in the energy range 
\begin{equation}\label{iE}
\frac{5.2\GeV}{(1+\z)^4+ \left(B/3.2\mkG\right)^2} = \frac{4D}{\beta R^2}\ll E \ll \frac{4D}{\beta h^2} = \frac{5.2\times10^4\GeV}{(1+\z)^4+ \left(B/3.2\mkG\right)^2}\,,
\end{equation}
the half-lifetime mean free path is
\begin{equation}\label{Li}
R/2\gg \Lambda_{1/2}\gg h/2\,.
\end{equation}
Electrons of these energies propagate diffusively out of the disc, to $|z|\gg h$, but travel along the radius over only modest distances. 
These are the particles which emit at $\nu=1.42$ and $0.408\GHz$. 
In this energy range,
\begin{equation}\label{NiE}
	N(\vec{x},E)\approx \frac{K}{\pi^{3/2}\beta E^2}\,\frac{\e^{-r^2/R^2}}{R^2}
	\int_E^\infty \dd E_0\,\frac{E_0^{-2}\e^{-z^2/(4\Lambda^2)}}{2\Lambda}.
\end{equation}

In terms of the integration variable $\xi=(1-E/E_0)^{1/2}$, the corresponding indefinite integral reduces to $\int \exp(-a^2/\xi^2)\,\dd\xi = \xi\exp(-a^2/\xi^2)+\sqrt{\pi} a\erf(a/\xi)+\const$, where $a=z/\sqrt{4D/(\beta E)}$ and $\erf(x)=(2/\sqrt{\pi})\int_0^x\exp(-t^2)\,\dd t$ with $\erf(\pm\infty)=\pm1$. Thus,
\begin{align}\label{NrE2}
	N(\vec{x},E)&\approx  \frac{K E^{-5/2}}{R^2 \sqrt{\pi^3\beta D}} \exp\left(-\frac{r^2}{R^2} \right)
	\left\{\exp\left(-z^2\frac{\beta E}{4D}\right)
	+ z\sqrt{\frac{\pi\beta E}{4D}} \left[\erf\left(z\sqrt{\frac{\beta E}{4D}} \right) -1 \right] \right\}.
\end{align}
The asymptotic spectral index at $z=0$ is smaller by 1/2 than that at the higher energies (Section~\ref{I1}).

\begin{figure}
	\centering
	\includegraphics[width=1.0\textwidth]{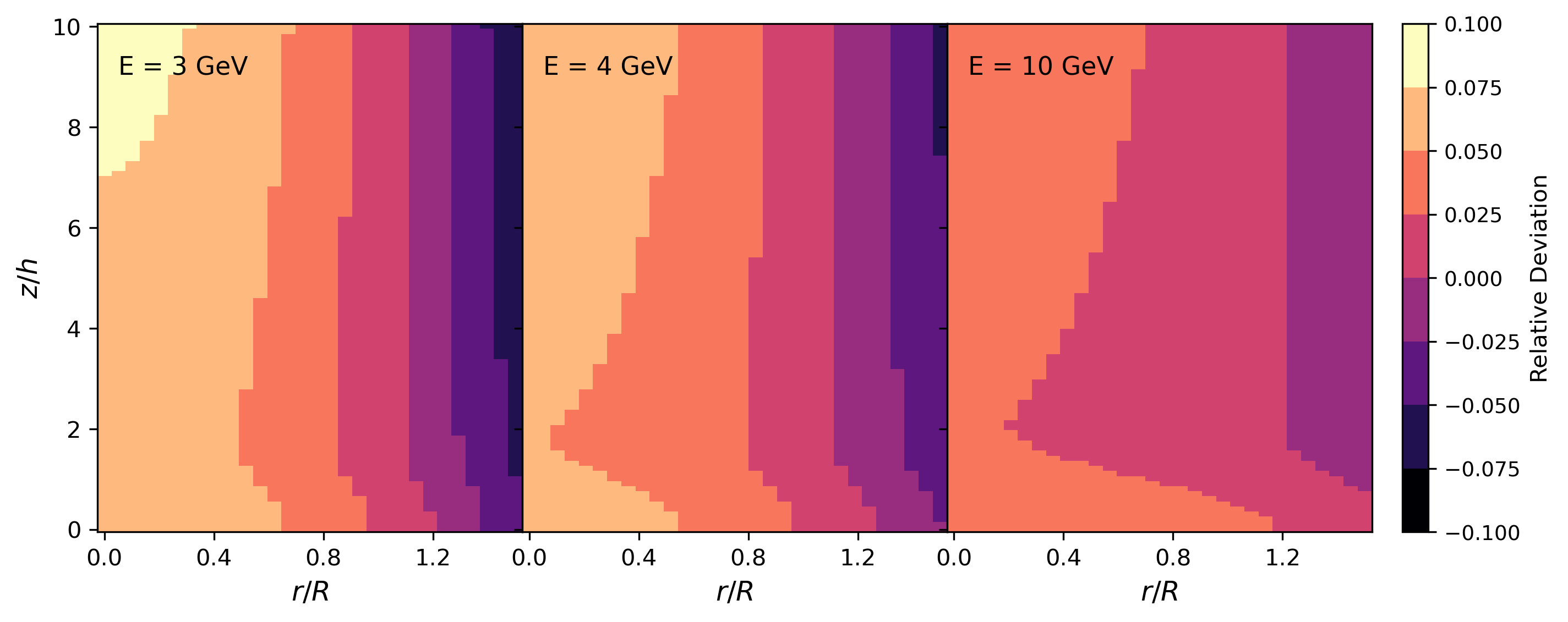}
	\caption[\ass{Ratio} \as{The relative deviations} of approximate \ass{to actual} CRE distributions at $E=3,4,10\GeV$ \as{from the exact values.}]{The \ass{ratio} \as{relative deviations} of  the approximate CRE spatial distributions $N(r,z,E)$ at $E = 3, 4$ and $10\GeV$ obtained using the refined approach of Section~\ref{app:mid_energy} \ass{to} \as{from} the exact values resulting from equation~\eqref{NQC}. \label{fig_ratio}}
    \end{figure}

In this energy range, 
$\exp\left(-z^2/(h^2+4\Lambda^2)\right)\approx \exp\left(-z^2/(4\Lambda^2)\right) \left[1+h^2z^2/(16\Lambda^4)\right]$, and the approximation for $N(\vec{x},E)$ is valid when the last term in the square brackets can be neglected, i.e., for
\begin{equation}\label{Zmax1}
	|z_{1/2}|\lesssim 4\Lambda_{1/2}^2/h\simeq h
\end{equation}
for the smallest value of $\Lambda_{1/2}$ in the range \eqref{Li} \as{(although it remains accurate at much larger values of $|z|$ -- see Fig.~\ref{fig_ratio})}. This constraint becomes less restrictive at lower energies as $\Lambda_{1/2}$ increases with decreasing $E$. The constraint for the radial range is, similarly to equation~\eqref{Zmax},
\begin{equation}\label{RmaxI}
	r_{1/2}\lesssim R^2/(2\Lambda_{1/2})\simeq R\,,
\end{equation}
for the largest  value of $\Lambda_{1/2}$ in the range \eqref{Li}.

\subsubsection{Low energies}\label{I3}

For low-energy particles, $4\Lambda^2\gg R^2$, i.e.,
\begin{equation}\label{lE}
	E \ll \frac{4D}{\beta R^2} \simeq \frac{5.2\GeV}{(1+\z)^4+ \left(B/3.2\mkG\right)^2} \,,
    \qquad
    \Lambda_{1/2}\gg R/2\,,    
\end{equation}
we have $R^2+4\Lambda^2 \approx 4\Lambda^2$, $h^2+4\Lambda^2 \approx 4\Lambda^2$, and the integral over $E$ of equation~\eqref{NQC} is evaluated using the dimensionless integration variable
\begin{equation}\label{xxi}
	\xi = \frac{4D}{R^2\beta}\left(\frac{1}{E}-\frac{1}{E_0} \right).
\end{equation}
The corresponding indefinite integral reduces to $\int \xi^{-3/2} \exp(-a^2/\xi)\,\dd\xi = -(\sqrt{\pi}/a)\erf(a/\sqrt{\xi})+\const$ with $a^2=(r^2+z^2)/R^2$. The particle distribution in this energy range follows as
\begin{equation}\label{NrE3}
	N(\vec{x},E)\approx \frac{K}{4\pi D \sqrt{r^2+z^2}} E^{-2}
    \left[1-\erf\sqrt{\frac{\beta E}{4D}(r^2+z^2)}\right].
\end{equation}

This approximation is valid when $|z_{1/2}|\ll 4\Lambda_{1/2}^2/h\simeq R^2/h$ and $r_{1/2}\ll 4\Lambda_{1/2}^2/R\simeq R$, but it is singular at $r^2+z^2 \to 0$ because $R^2$ and $h^2$ are neglected in the denominator of the integrand in equation~\eqref{NQC}. Therefore, the particle distribution near $\vec{x}=0$ has to be evaluated separately. 
For $\vec{x}=0$ and in terms of the variable \eqref{xxi}, equation~\eqref{NQC} reduces to
\begin{equation}\label{N0xi}
N(0,E)=\frac{K}{4\pi^{3/2}RDE^2} \int_0^{4D/(R^2\beta E)} 
\frac{\dd\xi}{(1+\xi)\sqrt{h^2/R^2 + \xi}}\,.
\end{equation}
\citet[][section 5.4]{BBGDP90} note that this integral is independent of $E$ for $E\ll 4D/(\beta R^2)$. For $h/R=10^{-2}$, it is approximately equal to 3.12 and weakly depends on $h/R$, being about 3.14 for $h/R=10^{-3}$, 2.96 for $h/R=0.1$ and 2 for $h/R=1$. Thus, a suitable expression for $N(0,E)$ in the low-energy range is given by
\begin{equation}\label{N0E}
 N(0,E)\approx\frac{3K}{4\pi^{3/2}RD}E^{-2}\,.
\end{equation}

Particles in the low-energy range lose energy slowly and thus propagate far from their sources. Therefore, their energy spectrum is asymptotically the same as the injection spectrum, $s=2=s_0$. In some parameter ranges (e.g., if $B\gg 5\mkG$), this energy range can also be important for the synchrotron emission at $\nu \leq 1.4\GHz$.

\begin{figure}
	\centering
	\includegraphics[width=0.5\textwidth]{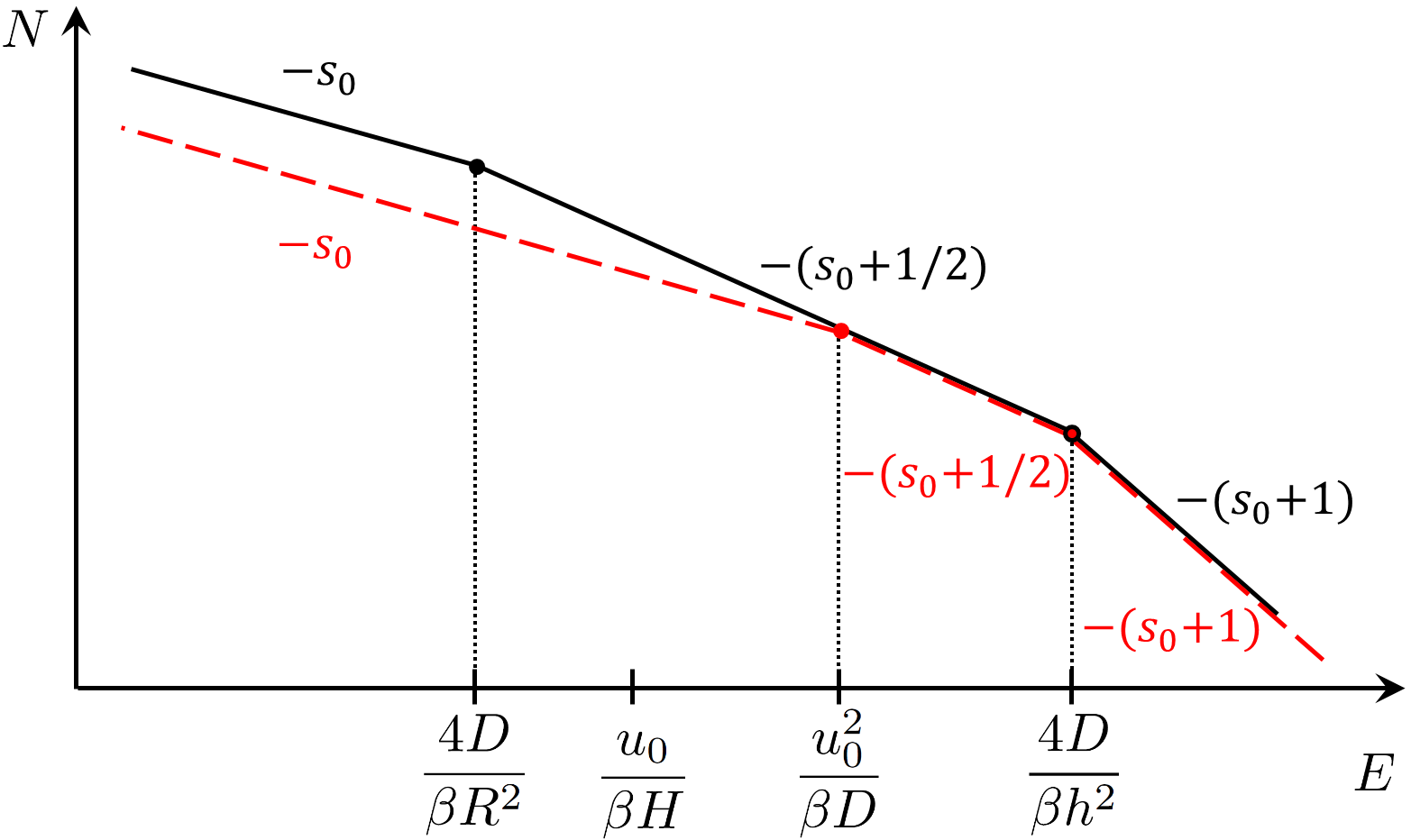}
	\caption[Electron energy spectra]{The energy spectra of relativistic electrons (in the double logarithmic scale) in the diffusion (solid/black) and diffusion--advection (dashed/red, outflow speed $u_0$ \as{and the vertical halo size $h$}) approximations \as{neglecting the dependence of the diffusivity on position and energy} \citep{DKP80}. The asymptotic power-law spectral indices are shown next to each part of the spectra, with $s_0$ the injection spectral index.
	}
	\label{ESS}
\end{figure}

\begin{figure}
	\centering
	\includegraphics[width=0.8\textwidth]{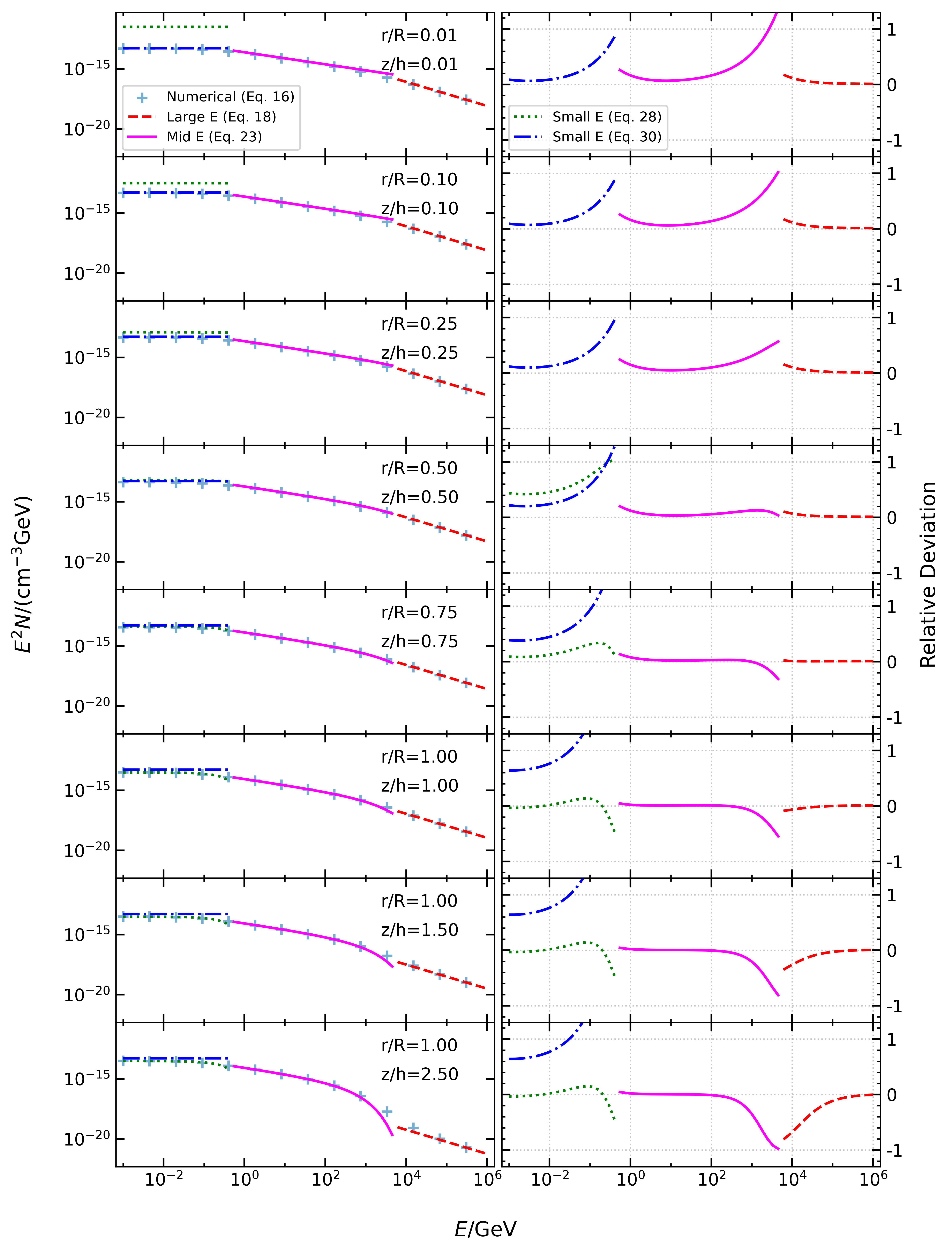}
	\caption[\as{Compensated} electron energy spectra at various locations]{The \as{compensated} energy spectra of relativistic electrons obtained from equation~\eqref{NQC} (symbols) for different values of $r$ and $z$ (indicated in the legends of the left-hand column) are compared with their corresponding approximations \eqref{NrE1} (red/dashed), \eqref{NrE2} (purple/solid), \eqref{NrE3} (green/dotted) and \eqref{N0E} (blue/dash-dotted). The right-hand panels show the \ass{ratio} \as{relative deviation} of  the approximate CRE energy spectra \ass{to} \as{from} the exact $N(\vec{x},E)$ \as{obtained} from equation~\eqref{NQC}. \label{ESSrz} 
}
\end{figure}

\begin{figure}
	\centering
	\includegraphics[width=0.8\textwidth]{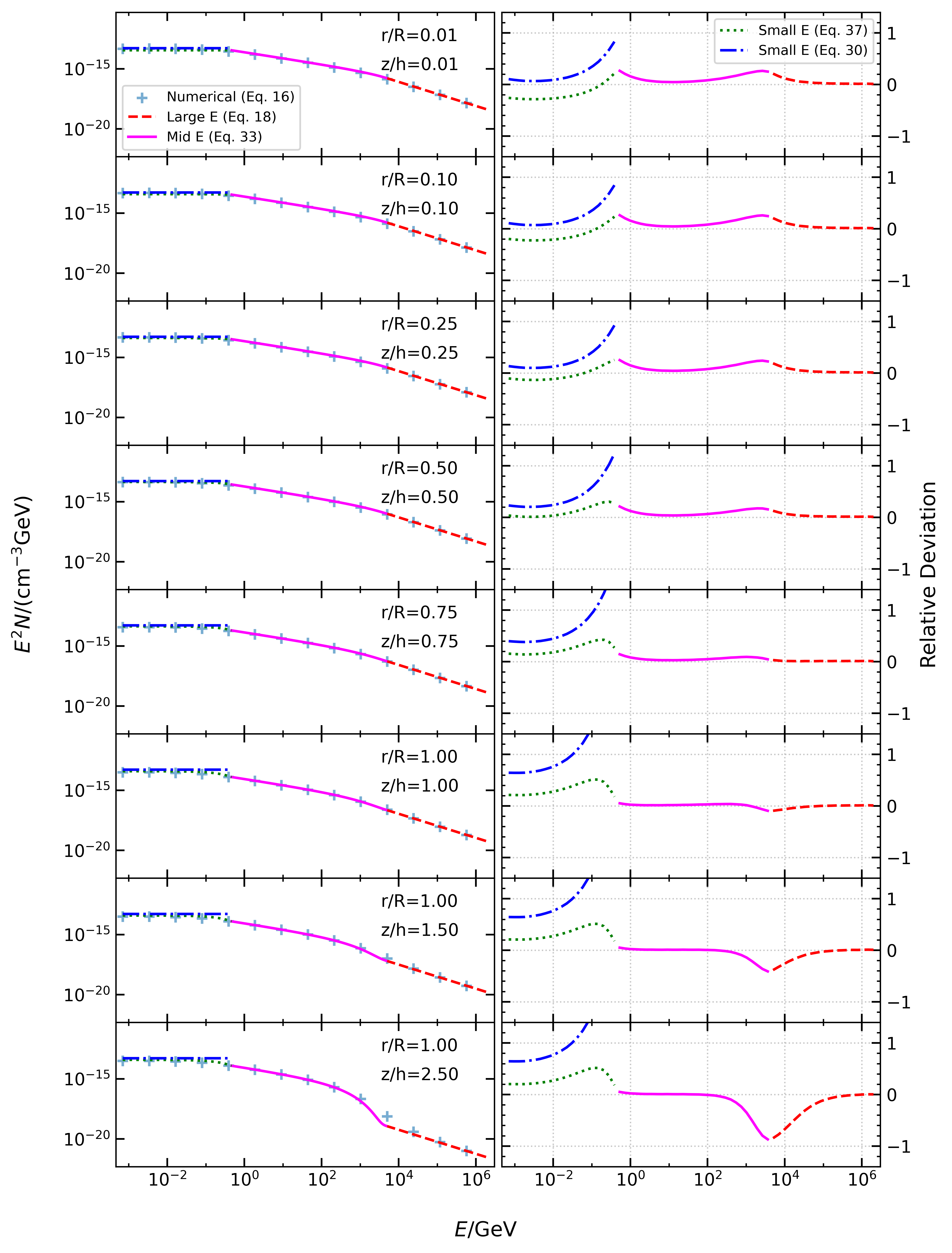}
	\caption[\as{Compensated} electron energy spectra at various locations \as{from the improved approximations}]{As Fig.~\ref{ESSrz} but for the improved approximations of Section~\ref{app:mid_energy}. \label{ESSrzi} 
	}
\end{figure}

\subsection{Quality of the approximations} \label{QoA}

A schematic form of the energy spectrum is shown in Fig.~\ref{ESS}. The energy spectra obtained from equation~\eqref{NQC}, where no approximations are involved, are compared with the approximate results \eqref{NrE1}, \eqref{NrE2}, \eqref{NrE3} and \eqref{N0E} at various values of $r$ and $|z|$ in Fig.~\ref{ESSrz}. The parameters used in Fig.~\ref{ESSrz} are: the redshift $\z =0$, $R = 10 \kpc$, $h = 0.1 \kpc$, $B = 10 \mkG$ and the star formation rate of $\dot M = 1 M_\odot/\yr$.

At high and intermediate energies ($4\Lambda^2 \gg h^2$ and $h^2 \ll 4\Lambda^2 \ll R^2$, respectively), the approximations given by equations~\eqref{NrE1} and \eqref{NrE2} accurately reproduce both the spatial distribution and the energy spectrum of CRE over all values of $r$ and $|z|$. The high-energy approximation remains accurate to within 10\% up to at least $r/R = 1$ and $|z|/h = 1$, even at the threshold energy $4\Lambda^2 = h^2$. The intermediate-energy approximation likewise achieves $\sim 10\%$ accuracy around $r/R \simeq 0.5$ and $|z|/h \simeq 0.5$. Equation~\eqref{NrE2}  becomes somewhat less accurate at smaller and especially larger distances from the origin still remanning quite reasonable near the middle of this energy range.

At low energies ($4\Lambda^2 \gg R^2$), the approximation given by equation~\eqref{NrE3} agrees well with the exact result of equation~\eqref{NQC} for $r/R \gtrsim 0.75$ and $|z|/h \gtrsim 0.75$, although some discrepancy develops near the transition scale $4\Lambda^2 = R^2$. Closer to the origin, however, the overall amplitude of the approximate solution becomes significantly larger than that of the exact solution as the approximation diverges at $r=z=0$, as discussed in Section~\ref{I3}. Equation~\eqref{N0E} reproduces the particle number density quite accurately for $r/R \leq 0.2$ and $|z|/h\leq 0.2$, whereas equation~\eqref{NrE3} can be used at larger distances from the origin. Since the diffusion length $\Lambda\gtrsim R/2$ is so large at the low energies and the boundary conditions are $\partial N /\partial r = 0$ at $r=0$ and $\partial N/\partial z = 0$ at $z=0$, it can be expected that $N(\vec{x},E)\approx N(0,E)$ in a wide region out to $r/R\simeq |z|/h\simeq 1$. 

\subsection{Refined approximations at intermediate and low energies} \label{app:mid_energy}

The accuracy of the approximations of Sections~\ref{I2} and \ref{I3} can be improved by considering more carefully the relation between the mean free path of the particles,
\begin{equation}\label{LEE0}
\Lambda(E,E_0) =\left[\frac{D}{\beta}
    \left(\frac{1}{E}-\frac{1}{E_0}\right)\right]^{1/2}\,,
\end{equation}
and $h/2$ and $R/2$ in the intermediate and low energy ranges.  Particles with $E_0\approx E$ travel over short distances $\Lambda(E,E_0)< h/2$ before they lose their energy irrespectively of their energy $E$. Their contribution to $N(\vec{x},E)$ is therefore similar to that of high-energy particles even when $E$ is in the intermediate energy range. Similarly, particles with $\Lambda(E,E_0)< h/2$, $h/2<\Lambda(E,E_0) < R/2$ and $\Lambda(E,E_0)>R/2$ are distributed differently in the low-energy range. These refinements are presented in this section and the results of a more consistent description across small, intermediate, and large diffusion scales are presented in Fig.~\ref{ESSrzi} for the energy spectra at various locations and in Fig.~\ref{fig_ratio} for the spatial distribution at selected energies.  

\subsubsection{Intermediate energies}\label{IEi}
The integral of equation~\eqref{NQC} in the energy range of Section~\ref{I2}, where $h/2\ll \Lambda_{1/2} \ll R/2$, still includes particles of initial energies $E_0$ which are close to $E$. Such particles travel over distances $\Lambda(E,E_0)$ shorter than $h/2$. Therefore, we introduce the energy $E_h$, such that $\Lambda(E,E_h)=h/2$:
\begin{equation}\label{Eh}
\frac{1}{E_h} =  \frac{1}{E} -\frac{h^2\beta}{4D}  \,,
\end{equation}
and split the integral of equation\eqref{NQC} into two energy ranges,
\begin{equation}\label{NrE2i}
N(\vec{x},E) = I_1+I_2\,,
\end{equation}
where the integral in $I_1$ extends over $E< E_0 < E_h$ and the integration range of $I_2$ is $E_h < E_0 < \infty$.
In the first integral, $0 < \Lambda(E,E_0) < h/2$ and these particles are better described with the approximation of Section~\ref{I1}. The calculation which leads to equation~\eqref{NrE1} but with the upper integration limit $E_h$ then leads to
\begin{equation} \label{NrE1-new}
     I_1 \approx \frac{K}{\pi^{3/2}\beta R^2h} E^{-2} \left(\dfrac{1}{E} -\dfrac{1}{ E_h} \right) \exp\left(-\frac{r^2}{R^2}-\frac{z^2}{h^2} \right). 
\end{equation} \label{Eq_appen1}
The second integral $I_2$ extends over the energies $E_0 > E_h$ where $\Lambda(E,E_0) > h/2$ and the approximation discussed in Section~\ref{I2} can be consistently applied with the lower integration limit $E_h$ rather than $E$:
\begin{align}
I_2 &\approx \frac{K}{R^2\sqrt{\pi^3\beta D}} E^{-5/2}\exp\left(-\frac{r^2}{R^2} \right)   \nonumber \\ 
      & \times \left\{ \exp\left(-\frac{z^2}{4\Lambda^2_\infty} \right) - 
      f_h \exp\left(-\frac{z^2}{4\Lambda_\infty^2 f_h^2}\right) +  
      \frac{z\sqrt{\pi}}{2\Lambda_\infty} \left[ \erf\left( \frac{z}{2\Lambda_\infty} \right) - 
        \erf\left( \frac{z}{2 \Lambda_\infty f_h} \right) \right] \right\}, 
\end{align} \label{Eq_appen2}
where 
\begin{equation}\label{fm}
f_h^2 = 1 - E/E_h\,,\quad 
\Lambda_\infty = \Lambda(E, \infty) =\sqrt{D/(\beta E)}\,.
\end{equation}

\subsubsection{Low energies}
The integration range of equation~\eqref{NQC} at low energies is similarly split into three energy intervals with
\begin{equation}\label{NrE3i}
N(\vec{x},E) = J_1+J_2+J_3\,.
\end{equation}
The integration in $J_1$ extends over $E < E_0 < E_h$: particles of these energies have $\Lambda(E,E_0) < h/2$. The energy range of $J_2$ is $E_h < E_0 < E_R$, where $h/2 < \Lambda(E,E_0) < R/2$ and
\begin{equation}\label{ER}
\frac{1}{E_R} = \frac{1}{E}-\frac{R^2\beta}{4D} \,.
\end{equation}
The energy range of $J_3$ is $E_R < E_0 <\infty$: these particles propagate over the largest distances, $\Lambda(E,E_0)>R/2$. 
Equation~\eqref{NrE1-new} is valid both for $I_1$ and $J_1$. 

The contribution $J_2$ can be evaluated using the approach of Section~\ref{I2} with the integration limits $E_h$ and $E_R$, leading to the expression similar to equation~\eqref{NiE}:
\begin{align}
J_2 &\approx  \frac{K }{R^2\sqrt{\pi^3\beta D}} E^{-5/2}\exp\left(-\frac{r^2}{R^2} \right)\nonumber\\
 	& \times\left\{  f_R\, \e^{-z^2/(4\Lambda_\infty^2 f_{R}^2)} 
    - f_{h}\, \e^{-z^2/(4\Lambda_\infty^2 f_{h}^2)}
    + \frac{z\sqrt{\pi}}{2\Lambda_\infty} 
    \left[ \erf\left( \frac{z}{2\Lambda_\infty f_R} \right) - \erf\left( \frac{z}{2 \Lambda_\infty f_h} \right) \right] \right\}, 
\end{align} \label{Eq_appen3}
with $f_{R}^2 = 1 -  E/E_R$. 

For the third contribution, $E_0 > E_R$ and $\Lambda(E,E_0)  > R/2$, so the low-energy (large-$\Lambda$) approximation of  Section~\ref{I3} applies. The result similar to equation~\eqref{NrE3} but obtained with the integration range $E_R< E_0<\infty$ has the form
\begin{equation}
J_3 \approx 
	 \frac{ K }{4\pi D \sqrt{r^2+z^2}} E^{-2}
    \left( \erf\sqrt{\frac{r^2+z^2}{\Lambda_\infty^2 f_{R}^2}} -\erf\sqrt{\frac{r^2+z^2}{\Lambda_\infty^2 }} \right),
\end{equation}
and we note that this expression is finite at $r^2+z^2=0$ unlike the low-energy approximation of equation~\eqref{NrE3}.

The approximate energy spectra obtained in this section are compared with the exact solution, equation~\eqref{NQC}, at various values of $r$ and $|z|$ in Fig.~\ref{ESSrzi}. Overall, the approximations show good agreement with the exact result with the root-mean-square errors in the range $10$--$20\%$ for $r/R \leq 1$ and $|z|/h \leq 1$ across the whole energy range considered, with the strongest deviations near the borderline energies between different approximations.

In Fig.~\ref{fig_ratio}, we  present the ratio of the approximate CRE distribution in space to the exact $N(\vec{x},E)$ from equation~\eqref{NQC} at three representative energies ($E = 3,\,4,$ and $10\GeV$), evaluated over a wider spatial domain $r/R \leq 1.5$ and $z/h \leq 10$. These energies approximately correspond to electrons radiating at \as{the} rest-frame frequencies of \ass{$1.42\,\GHz$, $408\,\MHz$, and $150\,\MHz$,} \as{$150\,\MHz$, $408\,\MHz$ and $1.42\,\GHz$,}  respectively, computed using equation~\eqref{Em}. Across this wide spatial range, the fractional differences between the approximate and exact solutions remain within $\pm8\%$.

\section{Arbitrary spatial distribution of cosmic ray sources}
\label{ARDCRS}
Since the propagation equation~\eqref{ADeq} is linear in $N$, the approximate solutions for $N(\vec{x},E)$ derived above can be used to derive the particle distributions for a wide range of the source distributions by superimposing a number of source functions of the form~\eqref{Q} if the boundary conditions are homogeneous, $\partial N/\partial r = 0$ at $r=0$, $\partial N/\partial z = 0$ at $z=0$, $N\to0$ at $|\vec{x}|\to\infty$ and $N\to0$ for $E\to\infty$. If the source is the superposition of several components, $Q(r,z)=\sum_k Q_k(r,z)$, the solution of equation~\eqref{ADeq} with homogeneous boundary conditions is also the superposition, $N(\vec{x},E)=\sum_k N_k(\vec{x},E)$, where $N_k(\vec{x},E)$ is the solution of equation~\eqref{ADeq} with $Q_k(r,z)$ on the right-hand side. Therefore, the results presented here can be applied to a variety of cosmic ray source distributions despite a rather specific form \eqref{Q} of the source $Q(r,z)$ for which they are obtained.

For example, the radial distribution of the number of pulsars per unit area in the Milky Way at $r\gtrsim0.5\kpc$ \citep{Lor04},
\begin{equation}\label{pul}
N_\text{p} = 64.6 \kpc^{-2}  \, (r/1\kpc)^{2.35}  \exp(-r/1.528\kpc)\,,
\end{equation}
is not monotonic, with a maximum at about $r=3.6\kpc$. This distribution is approximated by $Q(r,z)\propto A[q_1(r) + q_2(r)]$ (and Gaussian distributions in $z$) with
\begin{equation}\label{pul_appr}
A=178\kpc^{-2}\,,\quad 
q_1(r) = \exp[-(r/6.9\kpc)^2],\quad
q_2(r) = -\exp[-(r/2.2\kpc)^2]\,, 
\end{equation}
with the accuracy within 3--4\% for $r \leq 13\kpc$. 

In particular, an exponential disc, $Q\propto \exp(-r/R)$, can be accurately approximated in a finite range of $r/R$ by a superposition of Gaussian functions based on the discretisation of the Laplace transform,
\begin{equation}
e^{-r/R} = \frac{1}{R\sqrt{\pi}} \int_{0}^{\infty}  \frac{1}{\sqrt{s}}\, 
\exp\!\left(-\frac{s}{R^2}-\frac{r^{2}}{4s}\right)  \,\dd s\,.
\label{LG_rep}
\end{equation}
The integrand has a maximum at
\begin{equation}
s_*=\tfrac14 R^2\left(1+\sqrt{1+4r^2/R^2} \right).
\end{equation}
Equation~\eqref{LG_rep} can be discretised using finite increments of uniform length in $\ln s$, $\Delta = \Delta(\ln s) =\const$ (the corresponding increment of $s$ is $s\Delta$), as a sum centred at $s=s_*$:
\begin{equation} \label{LG_dis}
e^{-r/R}\approx \frac{\Delta}{R\sqrt{\pi}} \sum_{m=-M}^{M} \sqrt{s_m}\,  
                  \exp\!\left(-\frac{s_m}{R^{2}} \right) \exp \!\left(-\frac{r^{2}}{4s_m}\right),
\end{equation}
where $s_m=s_* \exp(m\Delta)$.

Expanding the integrand of equation~\eqref{LG_rep} to the second order in $\Delta$ about $s_\ast$ yields in equation~\eqref{LG_rep} a Gaussian integrand in $\ln s$ with the half-width
\begin{equation}
\sigma= \left( \frac{s_\ast}{R^{2}}+\frac{r^{2}}{4s_\ast} \right)^{-1/2}.
\end{equation}
In most applications, the sum of equation~\eqref{LG_dis} can be truncated to the interval 
$|\ln s- \ln s_*|\leq 5\sigma$. Using the discretisation interval $\Delta\simeq1.0$, we find excellent agreement between equations~\eqref{LG_rep} and \eqref{LG_dis} with the root-mean-square relative error of about $0.2\%$ and a maximum relative error below $1\%$ (at small $r$) over the range $0<r/R<2$. The number of terms required to achieve this level of accuracy varies from about $M=17$ at  $r \approx 0$ to $9$ at  $r/R \simeq 2$.

For an arbitrary distribution of the CR sources $Q\propto f(r/R)$, the parameters $w_i$ and $s_i$ of the approximation
\begin{equation} \label{Arb_source}
 f(r/R)  \approx \sum_{i=1}^{N} w_i\, 
    \exp\!\left(-\frac{r^{2}}{4s_i}\right)
\end{equation}
can be obtained from a least squares fit for a selected $N$ and the range of $r/R$ where the approximation is evaluated.

Similar approach can be used to include arbitrary distribution of the CRE sources in $z$. For the particle source distributions represented as a superposition of the Gaussians~\eqref{Q}, exact spatial distributions and energy spectra of CRE can be obtained using the corresponding sum of integrals in energy of the form~\eqref{NQC} in which the injection spectral index can be different from $s_0=2$ \as{and $D$ can be a function of the particle energy}. 

\as{This approach to include arbitrary distributions of the CRE sources requires that the form of the source distribution $Q$ is separable in $r$ and $z$, as in equation~\eqref{Q} (which is not the case when, e.g., the scale height $h$ depends on $r$)}.

\as{The spatial distributions of cosmic ray sources (star formation rate) convolved with Gaussian or exponential kernels have been successfully used in modelling the synchrotron emission of spiral galaxies \citep{VSBP20}. We stress, however, that the results presented here can be applied to an arbitrary distribution of the sources, including non-monotonic spatial distributions, provided they have a form factorizable in $r$ and $z$.}

\section{Discussion and conclusions}\label{DC}
We have derived explicit solutions of the diffusion equation for the propagation of relativistic electrons which are  sufficiently accurate (Sections~\ref{QoA} and \ref{app:mid_energy}) to be used in such problems as the calculation of the galactic radio luminosities for statistically large galaxy samples where solution of the propagation equations is computationally expensive. These solutions can also be useful in the interpretation of synchrotron observations of resolved galaxies as the only simple alternative to the assumption of equipartition between cosmic ray and magnetic fields (which is likely to be less accurate than the solutions presented here). 

The steady-state, \as{axially symmetric} spatial distribution and energy spectrum of relativistic electrons are obtained assuming that the particle diffusion is isotropic with the diffusivity $D$ independent of position and the particle energy.
\as{This assumption simplifies the calculations significantly. However, the cosmic ray diffusion coefficient is expected to vary with the particle energy at $E\gtrsim 5\GeV$ as $D\propto E^k$ with $k=1/3\text{--}1/2$ depending on the diffusion model \citep[see, e.g., section~1.5.1 of][for a review]{HSG21}. This would lead to steeper electron energy spectra \citep{BA12}. Equation~\eqref{NQC} remains applicable when the diffusivity $D$,  together with $\Lambda$, depends on the particle energy. At high energies discussed in Section~\ref{I1}, the diffusive mean free path of the particles is much smaller than the size of the source. As a result, the energy dependence of the diffusivity does not affect the particle distribution and equation~\eqref{NrE1} remains valid. At intermediate energies (Section~\ref{I2}), the integral over $E$ can no longer be expressed in terms of elementary functions for any $k\neq0$. The integral over $E$ at low energies, Section~\ref{I3} can be evaluated explicitly for $k=1/2$ but we do not present this rather cumbersome result. The cosmic ray diffusivity is also likely to increase with $|z|$ into galactic haloes. The spatial variation of $D$ is not included in equation~\eqref{NQC} and it is not clear if it can be included in the framework presented here.}

\label{AoA}
We also assume that the synchrotron and inverse Compton energy loss rate $\beta$ is independent of position. The variation of the magnetic field strength with $r$ and $z$ can be included by replacing the magnetic field strength $B$ in equation~\eqref{Eloss} with its root-mean-square value within a selected region, and $\Lambda_{1/2}=(D\tau_{1/2})^{1/2}$ or larger appears to be a natural choice for such a region size. 
We note in this connection that \citet{MFBMS16} find that the spectral index of the nonthermal radio emission in the galaxy M51 derived from observations at four frequencies in the range $\nu=151\MHz\text{--}1.4\GHz$ is hardly sensitive to variations of the galactic magnetic field with position \as{which are quite strong in their model.} The CRE propagation model used by these authors is strongly simplified as only the particle distribution along the galactocentric radius is solved for, with the diffusion along the vertical direction described as a loss term $-N/T$ in equation~\eqref{ADeq} with $T$ dependent on the diffusivity $D$. Nevertheless, their results suggest that a piece-wise constant approximation to the magnetic field strength \as{can} lead to reasonably accurate results.

Specific forms of $\beta$ used in the numerical values of the borderline energies, such as equations~\eqref{hE}, \eqref{iE}, \eqref{lE} and elsewhere, include the inverse Compton losses due to the CMB but neglect the contribution of the stellar radiation. The energy density of the stellar radiation in the Milky Way is  $w_\text{ph}\simeq0.7\eV\cm^{-3}\simeq10^{-12}\erg\cm^{-3}$ near the Sun  (table 12.1 of \citealt{Draine11}; section~2.3 of \citealt{S02}) but is higher by about an order of magnitude near the Galactic centre \citep{PJM17,PYTNRA17}. Galaxies with a higher star formation rate are likely to have still higher radiation densities. However, the CMB energy density  rapidly increases with the redshift and dominates over the stellar radiation in high-redshift galaxies with moderate star formation rates. Therefore, neglecting the energy losses to the stellar radiation field is an acceptable approximation for galaxies at higher redshifts, especially at redshifts $\z \gtrsim 2$ when the energy losses to the CMB photons dominate over the synchrotron losses  \citep{LT10,SB13,SSK15}. The effect of the spatially varying stellar radiation field on the electron density distribution can be included in the same manner as the spatial variations of the magnetic field.

The explicit forms  for the electron number density $N(\vec{x},E)$ are obtained in Sections~\ref{QoA} and \ref{app:mid_energy} for a specific form of the particle source \eqref{Q} where $Q\propto \exp(-r^2/R^2-z^2/h^2)$. However, we show in Section~\ref{ARDCRS} how our results can be used to derive $N(\vec{x},E)$ for an arbitrary spatial distribution of the particle sources since it can be represented as a superposition of the forms \eqref{Q}.
Overall, the explicit solutions and  approximations developed here provide a practical and physically motivated way to compute the cosmic-ray electron density from first principles without resorting to fully numerical propagation models. The demonstrated accuracy and the ability to treat arbitrary source distributions make this approach computationally efficient for applications to large galaxy samples, including synchrotron observations.


\as{Solutions of the electron transport equation presented here can be generalized to include the effects of advection (and streaming). Advection at a speed $u$ does not affect the particle distribution established by their diffusion as long as the advection time scale $z/u$ at an altitude $z$ is much longer than the diffusion time $z^2/D$, i.e., for $|z|\lesssim D/u\simeq 1\kpc\gg h$ for $u=100\kms$ \citep[see also \citealt{PVZB97}, and section~6 of][]{AB18}.  Thus, outflows only affect the spatial distribution of cosmic ray electrons at distances $|z|\gtrsim 1\kpc$ where both the magnetic field is weaker and the electron density is lower than at the mid-plane. Therefore, diffusion controls the distribution of cosmic ray electrons in many galaxies with a relatively weak outflow \citep[e.g.,][]{H+18,H2021}, including the Milky Way \citep{SMP07}.}

\as{\citet{DKP80} and \citet{BBGDP90} solve the one-dimensional problem in $z$ for the distribution of cosmic ray electrons in the diffusion--advection approximation with the advection speed independent of $z$ and the boundary condition $N|_{|z|=h}=0$, where $h$ is the vertical extent of the halo. These authors consider the energy range $E\ll D/(\beta h^2)$, so that the thickness of the source distribution is negligible in comparison with the particle path length $\Lambda$ and $Q(z,E)\propto \updelta(z)$ can be assumed. The resulting energy spectrum affected by the advection is shown in Fig.~\ref{ESS}. Green's function of the advection--diffusion equation with a constant advection speed can be derived, and the analysis presented here is likely to succeed in this case as well. Such a solution, applicable at large $|z|$ can then be matched with the diffusive solutions presented here which remain applicable closer to the mid-plane. A detailed treatment of the analytic solution using Green’s function for the constant-speed advection–diffusion equation is left for future work.}

\section*{Conflict of Interest Statement}
The authors declare that the research was conducted in the absence of any commercial or financial relationships that could be construed as a potential conflict of interest.

\section*{Author Contributions}
AS: Conceptualisation, Methodology, Formal analysis, Investigation, Writing – original draft, Writing – review \& editing; CJ: Methodology, Formal analysis, Investigation, Validation, Writing – original draft, Writing – review \& editing.

\section*{Funding}
CJ is supported by the Rashtriya Uchchatar Shiksha Abhiyan (RUSA) scheme (No.CUSAT/PL(UGC). A1/2314/2023, No:T3A). 

\section*{Acknowledgments}
We are grateful to \as{Rainer Beck,}  Luke Chamandy, Vladimir Dogiel, Sukanta Ghosh and Kandaswamy Subramanian for useful discussions and suggestions. \as{We gratefully acknowledge useful comments and suggestions of anonymous referees.}


\section*{Data Availability Statement}
The data used in this study are available in the text.

\bibliographystyle{Frontiers-Harvard}
\bibliography{ref}

@ARTICLE{AAV95,
       author = {{Atoyan}, A.~M. and {Aharonian}, F.~A. and {V{\"o}lk}, H.~J.},
        title = "{Electrons and positrons in the galactic cosmic rays}",
      journal = {\prd},
         year = 1995,
        month = sep,
       volume = {52},
       number = {6},
        pages = {3265-3275},
          doi = {10.1103/PhysRevD.52.3265},
       adsurl = {https://ui.adsabs.harvard.edu/abs/1995PhRvD..52.3265A}
}

@ARTICLE{AB18,
       author = {{Amato}, Elena and {Blasi}, Pasquale},
        title = "{Cosmic ray transport in the Galaxy: A review}",
      journal = {Adv. \ Space Res.},
         year = 2018,
        month = nov,
       volume = {62},
       number = {10},
        pages = {2731-2749},
          doi = {10.1016/j.asr.2017.04.019},
archivePrefix = {arXiv},
       eprint = {1704.05696},
 primaryClass = {astro-ph.HE},
       adsurl = {https://ui.adsabs.harvard.edu/abs/2018AdSpR..62.2731A}
}

@ARTICLE{Bell78b,
       author = {{Bell}, A.~R.},
        title = "{The acceleration of cosmic rays in shock fronts - II.}",
      journal = {\mnras},
         year = 1978,
        month = feb,
       volume = {182},
        pages = {443-455},
          doi = {10.1093/mnras/182.3.443},
       adsurl = {https://ui.adsabs.harvard.edu/abs/1978MNRAS.182..443B}
}

@BOOK{BBGDP90,
       author = {{Berezinski\u{\i}}, V.~S. and {Bulanov}, S.~V. and {Ginzburg}, V.~L. and {Dogiel}, V.~A. and {Ptuskin}, V.~S.},
        title = "{Astrophysics of Cosmic Rays}",
         year = 1990,
publisher = "{North-Holland}",
address = "{Amsterdam}",
       adsurl = {https://ui.adsabs.harvard.edu/abs/1984acr..book.....B}
}

@ARTICLE{BSY14,
       author = {{Blandford}, Roger and {Simeon}, Paul and {Yuan}, Yajie},
        title = "{Cosmic ray origins: an introduction}",
      journal = {Nuclear Phys. B Proc. Suppl.},
         year = 2014,
        month = nov,
       volume = {256},
        pages = {9-22},
          doi = {10.1016/j.nuclphysbps.2014.10.002},
archivePrefix = {arXiv},
       eprint = {1409.2589},
 primaryClass = {astro-ph.HE},
       adsurl = {https://ui.adsabs.harvard.edu/abs/2014NuPhS.256....9B}
}

@ARTICLE{B13,
       author = {{Blasi}, Pasquale},
        title = "{The origin of galactic cosmic rays}",
      journal = {\aapr},
         year = 2013,
        month = nov,
       volume = {21},
          eid = {70},
        pages = {70},
          doi = {10.1007/s00159-013-0070-7},
archivePrefix = {arXiv},
       eprint = {1311.7346},
 primaryClass = {astro-ph.HE},
       adsurl = {https://ui.adsabs.harvard.edu/abs/2013A&ARv..21...70B}
}

@ARTICLE{BA12,
       author = {{Blasi}, Pasquale and {Amato}, Elena},
        title = "{Diffusive propagation of cosmic rays from supernova remnants in the Galaxy. I: spectrum and chemical composition}",
      journal = {\jcap},
         year = 2012,
        month = jan,
       volume = {2012},
       number = {1},
          eid = {010},
        pages = {010},
          doi = {10.1088/1475-7516/2012/01/010},
archivePrefix = {arXiv},
       eprint = {1105.4521},
 primaryClass = {astro-ph.HE},
       adsurl = {https://ui.adsabs.harvard.edu/abs/2012JCAP...01..010B}
}

@ARTICLE{DKP80,
       author = {{Dogiel}, V.~A. and {Kovalenko}, V.~M. and {Prishchep}, V.~L.},
        title = "{Cosmic-ray electrons in the diffusion-convection model of particle propagation}",
      journal = {\sovastl},
         year = 1980,
        month = jun,
       volume = {6},
        pages = {366-368},
       adsurl = {https://ui.adsabs.harvard.edu/abs/1980SvAL....6..366D}
}

@book{Draine11,
	title={Physics of the Interstellar and Intergalactic Medium},
	author={{Draine}, B.~T.},
	isbn={9780691122144},
	lccn={2010028285},
	url={https://books.google.co.in/books?id=VWBRmQEACAAJ},
	year={2011},
	publisher={Princeton Univ.\ Press},
	address = {Princeton}
}

@article{E+08,
doi = {10.1088/1475-7516/2008/10/018},
url = {https://doi.org/10.1088/1475-7516/2008/10/018},
year = {2008},
month = {oct},
publisher = {},
volume = {2008},
number = {10},
pages = {018},
author = {Evoli, Carmelo and Gaggero, Daniele and Grasso, Dario and Maccione, Luca},
title = {Cosmic ray nuclei, antiprotons and gamma rays in the Galaxy: a new diffusion model},
journal = {J.\ Cosmology Astropart.\ Phys.}
}

@BOOK{GS64,
       author = {{Ginzburg}, V.~L. and {Syrovatskii}, S.~I.},
        title = "{The Origin of Cosmic Rays}",
         year = 1964,
publisher = {Pergamon},
address = {Oxford},
       adsurl = {https://ui.adsabs.harvard.edu/abs/1964ocr..book.....G}
}

@ARTICLE{Hansen+2024,
       author = {{Hansen}, Samuel P. and {Lagos}, Claudia D.~P. and {Bonato}, Matteo and {Cook}, Robin H.~W. and {Davies}, Luke J.~M. and {Delvecchio}, Ivan and {Tompkins}, Scott A.},
        title = "{Modelling the galaxy radio continuum from star formation and active galactic nuclei in the SHARK semi-analytic model}",
      journal = {\mnras},
         year = 2024,
        month = jun,
       volume = {531},
       number = {1},
        pages = {1971-1987},
          doi = {10.1093/mnras/stae1235},
archivePrefix = {arXiv},
       eprint = {2405.05586},
 primaryClass = {astro-ph.GA},
       adsurl = {https://ui.adsabs.harvard.edu/abs/2024MNRAS.531.1971H}
}

@ARTICLE{HSG21,
       author = {{Hanasz}, Micha{\l} and {Strong}, Andrew W. and {Girichidis}, Philipp},
        title = "{Simulations of cosmic ray propagation}",
      journal = {Living Reviews in Computational Astrophysics},
         year = 2021,
        month = dec,
       volume = {7},
       number = {1},
          eid = {2},
        pages = {2},
          doi = {10.1007/s41115-021-00011-1},
archivePrefix = {arXiv},
       eprint = {2106.08426},
 primaryClass = {astro-ph.HE},
       adsurl = {https://ui.adsabs.harvard.edu/abs/2021LRCA....7....2H}
}

@ARTICLE{H2021,
       author = {{Heesen}, Volker},
        title = "{The radio continuum perspective on cosmic-ray transport in external galaxies}",
      journal = {\apss},
         year = 2021,
        month = dec,
       volume = {366},
       number = {12},
          eid = {117},
        pages = {117},
          doi = {10.1007/s10509-021-04026-1},
archivePrefix = {arXiv},
       eprint = {2111.15439},
 primaryClass = {astro-ph.GA},
       adsurl = {https://ui.adsabs.harvard.edu/abs/2021Ap&SS.366..117H}
}

@ARTICLE{H+18,
       author = {{Heesen}, V. and {Krause}, M. and {Beck}, R. and {Adebahr}, B. and {Bomans}, D.~J. and {Carretti}, E. and {Dumke}, M. and {Heald}, G. and {Irwin}, J. and {Koribalski}, B.~S. and {Mulcahy}, D.~D. and {Westmeier}, T. and {Dettmar}, R.-J.},
        title = "{Radio haloes in nearby galaxies modelled with 1D cosmic ray transport using SPINNAKER}",
      journal = {\mnras},
         year = 2018,
        month = may,
       volume = {476},
       number = {1},
        pages = {158-183},
          doi = {10.1093/mnras/sty105},
archivePrefix = {arXiv},
       eprint = {1801.05211},
 primaryClass = {astro-ph.GA},
       adsurl = {https://ui.adsabs.harvard.edu/abs/2018MNRAS.476..158H}
}

@ARTICLE{H25,
       author = {{Hopkins}, Philip F.},
        title = "{Cosmic rays on galaxy scales: progress and pitfalls for CR--MHD dynamical models}",
      journal = {arXiv e-prints},
         year = 2025,
        month = sep,
          eid = {arXiv:2509.07104},
          doi = {10.48550/arXiv.2509.07104},
archivePrefix = {arXiv},
       eprint = {2509.07104},
 primaryClass = {astro-ph.GA},
       adsurl = {https://ui.adsabs.harvard.edu/abs/2025arXiv250907104H}
}

@ARTICLE{I+24,
       author = {{Irwin}, Judith and {Beck}, Rainer and {Cook}, Tanden and {Dettmar}, Ralf-J{\"u}rgen and {English}, Jayanne and {Heesen}, Volker and {Henriksen}, Richard and {Jiang}, Yan and {Li}, Jiang-Tao and {Lu}, Li-Yuan and {Mele}, Crystal and {M{\"u}ller}, Ancla and {Murphy}, Eric and {Porter}, Troy and {Rand}, Richard and {Skeggs}, Nathan and {Stein}, Michael and {Stein}, Yelena and {Stil}, Jeroen and {Strong}, Andrew and {Walterbos}, Rene and {Wang}, Q. Daniel and {Wiegert}, Theresa and {Yang}, Yang},
        title = "{CHANG-ES XXXI{\textemdash}A Decade of CHANG-ES: What we have learned from radio observations of edge-on galaxies}",
      journal = {Galaxies},
         year = 2024,
        month = may,
       volume = {12},
       number = {3},
          eid = {22},
        pages = {22},
          doi = {10.3390/galaxies12030022},
       adsurl = {https://ui.adsabs.harvard.edu/abs/2024Galax..12...22I}
}

@ARTICLE{JCSSRB24,
       author = {{Jose}, Charles and {Chamandy}, Luke and {Shukurov}, Anvar and {Subramanian}, Kandaswamy and {Rodrigues}, Luiz Felippe S. and {Baugh}, Carlton M.},
        title = "{Understanding the radio luminosity function of star-forming galaxies and its cosmological evolution}",
      journal = {\mnras},
         year = 2024,
        month = aug,
       volume = {532},
       number = {2},
        pages = {1504-1521},
          doi = {10.1093/mnras/stae1426},
archivePrefix = {arXiv},
       eprint = {2402.15099},
 primaryClass = {astro-ph.GA},
       adsurl = {https://ui.adsabs.harvard.edu/abs/2024MNRAS.532.1504J}
}

@ARTICLE{K01,
       author = {{Kroupa}, Pavel},
        title = "{On the variation of the initial mass function}",
      journal = {\mnras},
         year = 2001,
        month = apr,
       volume = {322},
       number = {2},
        pages = {231-246},
          doi = {10.1046/j.1365-8711.2001.04022.x},
archivePrefix = {arXiv},
       eprint = {astro-ph/0009005},
 primaryClass = {astro-ph},
       adsurl = {https://ui.adsabs.harvard.edu/abs/2001MNRAS.322..231K}
}

@INPROCEEDINGS{K08,
       author = {{Kroupa}, P.},
        title = "{The IMF of simple and composite populations}",
    booktitle = {Pathways Through an Eclectic Universe},
         year = 2008,
       editor = {{Knapen}, J.~H. and {Mahoney}, T.~J. and {Vazdekis}, A.},
        month = jun,
        pages = {p.~3-15},
          doi = {10.48550/arXiv.0708.1164},
archivePrefix = {arXiv},
       eprint = {0708.1164},
 primaryClass = {astro-ph},
       adsurl = {https://ui.adsabs.harvard.edu/abs/2008ASPC..390....3K}
}

@ARTICLE{LT10,
       author = {{Lacki}, Brian C. and {Thompson}, Todd A.},
        title = "{The physics of the far-infrared--radio correlation. II. Synchrotron emission as a star formation tracer in high-redshift galaxies}",
      journal = {\apj},
         year = 2010,
        month = jul,
       volume = {717},
       number = {1},
        pages = {196-208},
          doi = {10.1088/0004-637X/717/1/196},
archivePrefix = {arXiv},
       eprint = {0910.0478},
 primaryClass = {astro-ph.CO},
       adsurl = {https://ui.adsabs.harvard.edu/abs/2010ApJ...717..196L}
}

@BOOK{L94,
   author = {{Longair}, M.~S.},
    title = "{High Energy Astrophysics. Volume 2. Stars, the Galaxy and the Interstellar Medium.}",
booktitle = {High energy astrophysics.~Volume 2.~Stars, the Galaxy and the interstellar medium., by Longair, M.~S..~ Cambridge Univ.\  Press, Cambridge (UK), 1994, 410 p., ISBN 0-521-43439-4, Price {\pound} 45.00, US 69.95 (cloth).~ISBN 0-521-43584-6, Price {\pound} 16.95, US 34.95 (paper).},
publisher = {Cambridge Univ.\  Press},
address = {Cambridge},
     year = 1994,
   adsurl = {http://adsabs.harvard.edu/abs/1994hea2.book.....L}
}

@INPROCEEDINGS{Lor04,
       author = {{Lorimer}, D.~R.},
        title = "{The Galactic population and birth rate of radio pulsars}",
    booktitle = {Young Neutron Stars and Their Environments},
         year = 2004,
       editor = {{Camilo}, Fernando and {Gaensler}, Bryan M.},
       series = {IAU Symposium 218},
        month = jan,
        pages = {p.~105-112},
          doi = {10.48550/arXiv.astro-ph/0308501},
archivePrefix = {arXiv},
       eprint = {astro-ph/0308501},
 primaryClass = {astro-ph},
       adsurl = {https://ui.adsabs.harvard.edu/abs/2004IAUS..218..105L}
}

@ARTICLE{MFBMS16,
       author = {{Mulcahy}, D.~D. and {Fletcher}, A. and {Beck}, R. and {Mitra}, D. and {Scaife}, A.~M.~M.},
        title = "{Modelling the cosmic ray electron propagation in M51}",
      journal = {\aap},
         year = 2016,
        month = aug,
       volume = {592},
          eid = {A123},
        pages = {A123},
          doi = {10.1051/0004-6361/201628446},
archivePrefix = {arXiv},
       eprint = {1605.01406},
 primaryClass = {astro-ph.GA},
       adsurl = {https://ui.adsabs.harvard.edu/abs/2016A&A...592A.123M}
}

@ARTICLE{PYTNRA17,
       author = {{Popescu}, C.~C. and {Yang}, R. and {Tuffs}, R.~J. and {Natale}, G. and {Rushton}, M. and {Aharonian}, F.},
        title = "{A radiation transfer model for the Milky Way: I. Radiation fields and application to high-energy astrophysics}",
      journal = {\mnras},
         year = 2017,
        month = sep,
       volume = {470},
       number = {3},
        pages = {2539-2558},
          doi = {10.1093/mnras/stx1282},
       adsurl = {https://ui.adsabs.harvard.edu/abs/2017MNRAS.470.2539P}
}

@INPROCEEDINGS{PJM17,
       author = {{Porter}, T. and {Johannesson}, G. and {Moskalenko}, I.},
        title = "{The interstellar radiation field of the Milky Way in three spatial dimensions}",
    booktitle = {35th International Cosmic Ray Conference (ICRC2017)},
         year = 2017,
        month = jul,
          eid = {737},
        pages = {p.~737-744},
          doi = {10.22323/1.301.0737},
       adsurl = {https://ui.adsabs.harvard.edu/abs/2017ICRC...35..737P}
}

@ARTICLE{PVZB97,
       author = {{Ptuskin}, V.~S. and {V\"olk}, H.~J. and {Zirakashvili}, V.~N. and {Breitschwerdt}, D.},
        title = "{Transport of relativistic nucleons in a galactic wind driven by cosmic rays.}",
      journal = {\aap},
         year = 1997,
        month = may,
       volume = {321},
        pages = {434-443},
       adsurl = {https://ui.adsabs.harvard.edu/abs/1997A&A...321..434P}
}

@ARTICLE{R91,
       author = {{Rana}, Narayan C.},
        title = "{Chemical evolution of the Galaxy}",
      journal = {\araa},
         year = 1991,
        month = jan,
       volume = {29},
        pages = {129-162},
          doi = {10.1146/annurev.aa.29.090191.001021},
       adsurl = {https://ui.adsabs.harvard.edu/abs/1991ARA&A..29..129R}
}

@ARTICLE{RS19,
       author = {{Rephaeli}, Yoel and {Sadeh}, Sharon},
        title = "{Energetic particles in haloes of star forming galaxies}",
      journal = {\mnras},
         year = 2019,
        month = jun,
       volume = {486},
       number = {2},
        pages = {2496-2506},
          doi = {10.1093/mnras/stz963},
archivePrefix = {arXiv},
       eprint = {1904.01997},
 primaryClass = {astro-ph.HE},
       adsurl = {https://ui.adsabs.harvard.edu/abs/2019MNRAS.486.2496R}
}

@ARTICLE{RS24,
       author = {{Rephaeli}, Yoel and {Sadeh}, Sharon},
        title = "{Energetic particles in the central starburst, disc, and halo of NGC~253}",
      journal = {\mnras},
         year = 2024,
        month = feb,
       volume = {528},
       number = {2},
        pages = {1596-1603},
          doi = {10.1093/mnras/stae138},
archivePrefix = {arXiv},
       eprint = {2402.00523},
 primaryClass = {astro-ph.HE},
       adsurl = {https://ui.adsabs.harvard.edu/abs/2024MNRAS.528.1596R}
}

@ARTICLE{SB13,
       author = {{Schleicher}, Dominik R.~G. and {Beck}, Rainer},
        title = "{A new interpretation of the far-infrared--radio correlation and the expected breakdown at high redshift}",
      journal = {\aap},
         year = 2013,
        month = aug,
       volume = {556},
          eid = {A142},
        pages = {A142},
          doi = {10.1051/0004-6361/201321707},
archivePrefix = {arXiv},
       eprint = {1306.6652},
 primaryClass = {astro-ph.CO},
       adsurl = {https://ui.adsabs.harvard.edu/abs/2013A&A...556A.142S}
}

@BOOK{S02,
	author = {{Schlickeiser}, R.},
	title = "{Cosmic Ray Astrophysics}",
	booktitle = {Cosmic ray astrophysics / Reinhard Schlickeiser, Astronomy and 
	Astrophysics Library; Physics and Astronomy Online Library.~Berlin: 
	Springer.~ISBN 3-540-66465-3, 2002, XV + 519 pp.},
	publisher = {Springer},
	address = {Berlin},
	year = 2002,
	adsurl = {http://adsabs.harvard.edu/abs/2002cra..book.....S}
}

@ARTICLE{Schober+2023,
       author = {{Schober}, J. and {Sargent}, M.~T. and {Klessen}, R.~S. and {Schleicher}, D.~R.~G.},
        title = "{A model for the infrared-radio correlation of main sequence galaxies at gigahertz frequencies and its variation with redshift and stellar mass}",
      journal = {\aap},
         year = 2023,
        month = nov,
       volume = {679},
          eid = {A47},
        pages = {A47},
          doi = {10.1051/0004-6361/202245218},
archivePrefix = {arXiv},
       eprint = {2210.07919},
 primaryClass = {astro-ph.GA},
       adsurl = {https://ui.adsabs.harvard.edu/abs/2023A&A...679A..47S}
}

@ARTICLE{SSK15,
       author = {{Schober}, J. and {Schleicher}, D.~R.~G. and {Klessen}, R.~S.},
        title = "{X-ray emission from star-forming galaxies -- signatures of cosmic rays and magnetic fields}",
      journal = {\mnras},
         year = 2015,
        month = jan,
       volume = {446},
       number = {1},
        pages = {2-17},
          doi = {10.1093/mnras/stu1999},
archivePrefix = {arXiv},
       eprint = {1404.2578},
 primaryClass = {astro-ph.GA},
       adsurl = {https://ui.adsabs.harvard.edu/abs/2015MNRAS.446....2S}
}

@ARTICLE{S77,
       author = {{Segalovitz}, A.},
        title = "{The spectral index distribution of M51}",
      journal = {\aap},
         year = 1977,
        month = oct,
       volume = {61},
        pages = {59-67},
       adsurl = {https://ui.adsabs.harvard.edu/abs/1977A&A....61...59S}
}

@ARTICLE{SB19,
       author = {{Seta}, Amit and {Beck}, Rainer},
        title = "{Revisiting the equipartition assumption in star-forming galaxies}",
      journal = {Galaxies},
         year = 2019,
        month = apr,
       volume = {7},
       number = {2},
          eid = {45},
        pages = {45},
          doi = {10.3390/galaxies7020045},
archivePrefix = {arXiv},
       eprint = {1903.11856},
 primaryClass = {astro-ph.GA},
       adsurl = {https://ui.adsabs.harvard.edu/abs/2019Galax...7...45S}
}

@ARTICLE{S+2023,
       author = {{Stein}, M. and {Heesen}, V. and {Dettmar}, R.-J. and {Stein}, Y. and {Br{\"u}ggen}, M. and {Beck}, R. and {Adebahr}, B. and {Wiegert}, T. and {Vargas}, C.~J. and {Bomans}, D.~J. and {Li}, J. and {English}, J. and {Chy{\.z}y}, K.~T. and {Paladino}, R. and {Tabatabaei}, F.~S. and {Strong}, A.},
        title = "{CHANG-ES. XXVI. Insights into cosmic-ray transport from radio halos in edge-on galaxies}",
      journal = {\aap},
         year = 2023,
        month = feb,
       volume = {670},
          eid = {A158},
        pages = {A158},
          doi = {10.1051/0004-6361/202243906},
archivePrefix = {arXiv},
       eprint = {2210.07709},
 primaryClass = {astro-ph.GA},
       adsurl = {https://ui.adsabs.harvard.edu/abs/2023A&A...670A.158S}
}

@ARTICLE{SMP07,
       author = {{Strong}, Andrew W. and {Moskalenko}, Igor V. and {Ptuskin}, Vladimir S.},
        title = "{Cosmic-ray propagation and interactions in the Galaxy}",
      journal = {Ann.\ Rev.\ Nucl.\ Particle Sci.},
         year = 2007,
        month = nov,
       volume = {57},
       number = {1},
        pages = {285-327},
          doi = {10.1146/annurev.nucl.57.090506.123011},
archivePrefix = {arXiv},
       eprint = {astro-ph/0701517},
 primaryClass = {astro-ph},
       adsurl = {https://ui.adsabs.harvard.edu/abs/2007ARNPS..57..285S}
}

@ARTICLE{S+10,
       author = {{Strong}, A.~W. and {Porter}, T.~A. and {Digel}, S.~W. and {J{\'o}hannesson}, G. and {Martin}, P. and {Moskalenko}, I.~V. and {Murphy}, E.~J. and {Orlando}, E.},
        title = "{Global cosmic-ray-related luminosity and energy budget of the Milky Way}",
      journal = {\apjl},
         year = 2010,
        month = oct,
       volume = {722},
       number = {1},
        pages = {L58-L63},
          doi = {10.1088/2041-8205/722/1/L58},
archivePrefix = {arXiv},
       eprint = {1008.4330},
 primaryClass = {astro-ph.HE},
       adsurl = {https://ui.adsabs.harvard.edu/abs/2010ApJ...722L..58S}
}

@ARTICLE{S59,
       author = {{Syrovatskii}, S.~I.},
        title = "{The distribution of relativistic electrons in the Galaxy and the spectrum of synchrotron radio emission}",
      journal = {\sovast},
         year = 1959,
        month = feb,
       volume = {3},
        pages = {22},
       adsurl = {https://ui.adsabs.harvard.edu/abs/1959SvA.....3...22S}
}

@ARTICLE{Th+26,
       author = {{Thykkathu}, Nijin J. and {Jarvis}, Matt J. and {Whittam}, Imogen H. and {Hale}, C.~L. and {Matthews}, A.~M. and {Heywood}, I. and {Malefahlo}, Eliab and {Varadaraj}, R.~G. and {Stylianou}, N. and {Pearson}, Chris and {Seymour}, Nick and {Vaccari}, Mattia},
        title = "{MIGHTEE: The evolving radio luminosity functions of star-forming galaxies to $z\sim 4.5$ and the cosmic history of star formation}",
      journal = {\mnras},
         year = 2026,
        month = mar,
          doi = {10.1093/mnras/stag616},
archivePrefix = {arXiv},
       eprint = {2601.14913},
 primaryClass = {astro-ph.GA},
       adsurl = {https://ui.adsabs.harvard.edu/abs/2026MNRAS.tmp..573T}
}

@ARTICLE{VSBP20,
       author = {{Vollmer}, B. and {Soida}, M. and {Beck}, R. and {Powalka}, M.},
        title = "{Deciphering the radio star formation correlation on kpc scales. I. Adaptive kernel smoothing experiments}",
      journal = {\aap},
         year = 2020,
        month = jan,
       volume = {633},
          eid = {A144},
        pages = {A144},
          doi = {10.1051/0004-6361/201935923},
archivePrefix = {arXiv},
       eprint = {1911.02253},
 primaryClass = {astro-ph.GA},
       adsurl = {https://ui.adsabs.harvard.edu/abs/2020A&A...633A.144V}
}

@ARTICLE{Yoon2024,
       author = {{Yoon}, Ilsang},
        title = "{A simple model of the radio{\textendash}infrared correlation depending on gas surface density and redshift}",
      journal = {\apj},
         year = 2024,
        month = nov,
       volume = {975},
       number = {1},
          eid = {15},
        pages = {15},
          doi = {10.3847/1538-4357/ad7385},
archivePrefix = {arXiv},
       eprint = {2408.13469},
 primaryClass = {astro-ph.GA},
       adsurl = {https://ui.adsabs.harvard.edu/abs/2024ApJ...975...15Y}
}

\end{document}